# Stark deceleration of NO radicals

Xingan Wang[1], Moritz Kirste[1], Gerard Meijer[1,2], and Sebastiaan Y. T. van de Meerakker[2,1*]

1. Fritz-Haber-Institut der Max-Planck-Gesellschaft, Faradayweg 4-6, D-14195, Berlin, Germany

2. Radboud University Nijmegen, Institute for Molecules and Materials, Heijendaalseweg 135, 6525 AJ Nijmegen, the Netherlands

**Abstract:**
We report on the Stark deceleration of a pulsed molecular beam of NO radicals. Stark deceleration of this chemically important species has long been considered unfeasible due to its small electric dipole moment of 0.16 D. We prepared the NO radicals in the $X\ ^2\Pi_{3/2}$, v=0, J=3/2 spin-orbit excited state from the $X\ ^2\Pi_{1/2}$, v=0, J=1/2 ground state by Franck-Condon pumping via the $A\ ^2\Sigma^+$ state. The larger effective dipole moment in the J=3/2 level of the $X\ ^2\Pi_{3/2}$, v=0 state, in combination with a 316-stages-long Stark decelerator, allowed us to decelerate NO radicals from 315.0 m/s to 229.2 m/s, thus removing 47 % of their kinetic energy. The measured time-of-flight profiles of the NO radicals exiting the decelerator show good agreement with the outcome of numerical trajectory simulations.

**Introduction:**
Over the last decade, there has been a growing interest in the development of methods to manipulate the velocity of neutral molecules in a molecular beam. Using accelerators/decelerators that are the analogues for neutral molecules of LINACs (linear accelerators) for charged particles, full control over the velocity and velocity distribution of molecules can be obtained [1]. Interest in this approach is triggered by the various applications of these tamed molecular beams. Molecules decelerated to a near standstill, for instance, can be loaded and confined in traps, offering observation times that are orders of magnitude larger than what can be achieved in a beam [2-3]. Thus far, this has been exploited to measure the lifetimes of long-lived excited states, or to study collision processes at (ultra)low temperatures. The advantages of increased interaction times afforded by slowly moving molecules have also been exploited in high-resolution spectroscopy and metrology [4-6]. In crossed molecular beam experiments, these beams offer the possibility to study molecular collision processes as a function of the collision energy with a high intrinsic energy resolution [7-11].

A variety of approaches have been followed to accelerate or decelerate neutral molecules. In 1999, the first deceleration of molecules was demonstrated using an array of electric field electrodes that were switched to high voltages at the appropriate times. In this so-called Stark decelerator, the quantum-state specific force that polar molecules experience in an electric field is exploited. In this seminal experiment, a small part of a pulsed beam of metastable CO molecules was decelerated from 225 m/s to 98 m/s [12]. Since then, a variety of Stark decelerators have been designed and built, ranging in size from decelerators integrated on a chip to several meter long ones [13, 14]. Various species, including $ND_3$, OH, and NH, have been Stark-decelerated, trapped, and used for further experiments exploiting the exquisite properties

* Correspondence author (e-mail: b.vandemeerakker@science.ru.nl)

of Stark-decelerated molecular beams [1, 2, 15].

Inspired in part by these successes, a magnetic analogue of the Stark decelerator has been developed as well [16]. Deceleration based on the magnetic interaction allows the manipulation of a wide range of atoms and molecules to which the Stark deceleration technique cannot be applied. Zeeman deceleration of H, D, Ar and Ne atoms as well as $O_2$ molecules has been demonstrated recently [17-20]. The development of both the Stark and Zeeman deceleration techniques has been one of the most important advances in molecular beam technology in the last decade, and has been (and beyond doubt will be) used to advantage in a broad variety of research fields in molecular physics and physical chemistry [1, 21].

The Stark and Zeeman deceleration techniques can be applied to a relatively large variety of molecular species. The suitability of a molecule depends mostly on its mass and its effective electric or magnetic dipole moment. The choice for using a certain molecule further depends on the scientific relevance of the molecular species, as well as on experimental factors such as ease of production and detection. To date, the molecules $ND_3$, CO ($a^3\Pi$) and OH ($X\,^2\Pi$) have been used most successfully.

In view of many of these desiderata, the nitric oxide (NO) radical appears a prominent candidate for the Stark or Zeeman deceleration techniques. The NO radical is one of the most extensively studied molecular species to date, and plays a key role in atmospheric chemistry as well as in combustion and biological processes, and is an important intermediate in the chemical industry. Its significance to the scientific community was further cemented by its proclamation as "Molecule of the Year" in 1992. In molecular beam laboratories, the NO radical in the $X\,^2\Pi_{1/2}$ ground state has been one of the most favorite species to use in, for instance, scattering experiments [22-24]. NO is one of the few open-shell radicals that can be stored at large pressure in the gas-phase, and can be directly seeded at large quantities in a carrier gas. Extremely sensitive and easy to implement state-selective detection methods exist, using either Laser Induced Fluorescence (LIF) or Resonance Enhanced Multi Photon Ionization (REMPI) techniques.

The manipulation of NO radicals with electric or magnetic fields, however, is notoriously difficult. NO radicals in their $X\,^2\Pi_{1/2}$, $J=1/2$ rotational ground state possess only a small effective electric dipole moment of 0.05 Debye [25], whereas their magnetic dipole moment vanishes almost completely. This important species has therefore been exempt from the Stark or Zeeman deceleration techniques; the velocity of a beam of NO radicals could thus far only be manipulated using optical Stark deceleration and by billiard-like inelastic scattering processes [26,27].

Here, we report on the successful Stark deceleration of a pulsed molecular beam of NO radicals. Prior to Stark deceleration, a packet of NO radicals is prepared in the $X\,^2\Pi_{3/2}$, v=0, J=3/2 spin-orbit excited state from the $X\,^2\Pi_{1/2}$, v=0, J=1/2 ground state by Franck-Condon pumping via the $A\,^2\Sigma^+$ state. The factor 1.8 larger effective dipole moment of the J=3/2 level in the $X\,^2\Pi_{3/2}$, v=0 state allowed us to decelerate NO radicals from 315.0 m/s to 229.2 m/s using a 316-stages-long Stark decelerator, thus removing 47 % of their kinetic energy. The measured time-of-flight profiles of the NO radicals exiting the decelerator show good agreement with the outcome of numerical trajectory simulations. These velocity controlled beams of NO are ideal to perform high-resolution state-selective scattering experiments involving NO radicals and atoms, molecules, or surfaces.

**Experiment**

The experiments are performed in a Stark deceleration molecular beam apparatus that is schematically shown in Figure 1. A molecular beam of NO radicals is generated by expanding a gas mixture (20% NO seeded in Xenon, 3 bar) into vacuum using a pulsed valve (General valve series 99, 10 Hz repetition rate, 1-mm diameter orifice). The initial velocity of the molecular beam is reduced to approximately 315.0 m/s by cooling the valve body to -70 $^o$C. After the supersonic expansion, most NO radicals reside in the $X\ ^2\Pi_{1/2}$, v=0, J=1/2 rotational ground state. This rotational state consists of two Λ-doublet components of opposite parity. The lower and upper components of $e$ and $f$ parity, respectively, are separated by only 318 MHz and are equally populated in the molecular beam.

Before passing through a 2 mm diameter skimmer, part of the population in the $X\ ^2\Pi_{1/2}$, v=0, J=1/2, $e$ level is transferred to the spin-orbit excited $X\ ^2\Pi_{3/2}$, v=0, J=3/2, $f$ level by Franck-Condon pumping following electronic excitation to the $A\ ^2\Sigma^+$, N=1, J=1/2 and 3/2 negative-parity levels as shown in Figure 2. For this, the output of a pulsed dye laser, tuned to a wavelength of 226.2 nm, intersects the molecular beam at right angles approximately 10 mm downstream from the nozzle. The lifetime of the $A^2\Sigma^+$, N=1, J=1/2 and 3/2 levels is about 200 ns [28] and following optical excitation, the NO radicals spontaneously decay back to various vibrational levels in the $^2\Pi$ electronic ground state. To optimize the laser excitation, a photo multiplier tube (PMT) is mounted in the source chamber to monitor the fluorescence intensity. Sufficient laser intensity has been used to saturate the transition. Following the selection rules for electric dipole transitions, and the Hönl-London and Franck-Condon factors that govern ro-vibrational transitions, only about 6 % of the NO ($X\ ^2\Pi_{1/2}$, v=0, J=1/2, $e$) radicals are optically transferred into the desired $X\ ^2\Pi_{3/2}$, v=0, J=3/2, $f$ level. It is noted that this fraction can in principle be enhanced significantly when a second pulsed dye laser is added to allow for stimulated emission pumping, but this was not done in the present experiments.

The Stark shift of the $X\ ^2\Pi_{3/2}$, v=0, J=3/2 state is shown in Figure 2. The Λ-doublet splitting between the upper and lower components that are of $f$ and $e$ parity, respectively, is only 39 MHz, and is not visible on the energy scale of the figure. In the presence of an electric field, the upper Λ-doublet component splits into a $M_J\Omega$=-3/4 and $M_J\Omega$=-9/4 component, whereas the lower Λ-doublet component splits into a $M_J\Omega$=+3/4 and $M_J\Omega$=+9/4 component. For comparison, the Stark shift that is observed for NO radicals in the $X\ ^2\Pi_{1/2}$, v=0, J=1/2 is shown in Figure 2 as well. It can be seen that the effective dipole moment, i.e. the slope of the Stark shift, is significantly increased for NO radicals that are excited to the $X\ ^2\Pi_{3/2}$, v=0, J=3/2 state. Only molecules in the low-field-seeking $X\ ^2\Pi_{3/2}$, v=0, J=3/2, $f$, $M_J\Omega$=-9/4 component are decelerated in the experiments, although the $M_J\Omega$=-3/4 component also contributes to the measured time-of-flight profiles (*vide infra*). Approximately 25 mm from the nozzle orifice, the beam passes through a 2 mm diameter skimmer, and enters the decelerator chamber. Details of the Stark decelerator are described elsewhere [14] and we only give a brief description here. The 2.6-meter-long Stark decelerator consists of 316 electric field stages with a center-to-center distance of 8.25 mm. Each stage consists of two parallel 4.5 mm diameter electrodes that are placed orthogonally around the molecular beam axis to provide a 3 x 3 mm$^2$ area for the molecular beam to pass. A voltage difference of 40 kV is applied to the electrodes. A sequence of high voltage pulses is applied to the electrodes at the appropriate times to manipulate the velocity (and velocity distribution) of the packet of NO ($X\ ^2\Pi_{3/2}$, v=0, J=3/2, f) radicals.

The Stark decelerator can be used with different modes of operation, which are characterized by the parameters $\phi_0$ and s. The phase angle $\phi_0$ determines the amount of energy that is taken out from the molecules in every electric field stage, where $\phi_0$ = 0° corresponds to guiding the packet of molecules at constant velocity through the decelerator, and $\phi_0$ = 90° corresponds to maximum energy loss per stage. The parameter s determines the number of electrode pairs the molecular packet passes before the fields are switched [29]. Operation of a Stark decelerator with s > 1 can be advantageous, as these modes of operation decouple the longitudinal and transverse motions of the molecules inside the decelerator, eliminating loss of molecules during their flight through the decelerator. In general, for a Stark decelerator with given length, the highest possible deceleration (and thus the lowest final velocity) is obtained when operating the decelerator at high phase angles and s = 1. The highest possible intensity of the NO packet that exits the decelerator, however, is obtained when operating the decelerator at low phase angles and s = 3. In the experiments reported here, both the s = 1 and the s = 3 operation modes are used.

The NO $X\ ^2\Pi_{3/2}$, v=0, J=3/2 radicals that exit the decelerator are state-selectively detected using a Laser Induced Fluorescence scheme. The molecules are transferred to the $A\ ^2\Sigma^+$ state by inducing the Q$_{12}$ (3/2) and P$_2$(3/2) [$\Delta J_{F'F''}$(J'')] lines of the $A\ ^2\Sigma^+$, v=0 <- $X\ ^2\Pi$, v=0 transition around 227 nm. The transitions are induced under saturated conditions. The off-resonant fluorescence is collected by a lens and imaged onto a PMT. Stray light from the laser is suppressed by light baffles and by optical filters in front of the PMT.

**Results and Discussion**
In Figure 3, time-of-flight (TOF) profiles of NO ($X\ ^2\Pi_{3/2}$, v=0, J=3/2, f) radicals exiting the decelerator are shown for different modes of operation of the Stark decelerator. In curves (a) and (b) the TOF profile is shown that is obtained when the decelerator is operated at $\phi_0$ = 0° and s = 1 and s = 3, respectively. These modes of operation correspond to guiding a packet of NO at a constant velocity through the decelerator. A packet of NO radicals with a mean velocity of 315 m/s is selected, transported through the 2.6-meter-long decelerator, and arrives in the detection region some 8.5 ms after optical excitation in the source region. The wings of the TOF profile contain sharp features, as has been observed and interpreted before in deceleration experiments of OH radicals [29]. The signal intensity is about a factor two larger when the Stark decelerator is operated in the s = 3 mode compared to the s =1 mode, in agreement with earlier findings [14, 30].

In view of the relatively small Stark shift of NO ($X\ ^2\Pi_{3/2}$, v=0, J=3/2, f) radicals, deceleration is only possible when the Stark decelerator is operated in the s = 1 mode. In curve (c) of Figure 3, the

TOF profile is shown when the decelerator is operated at $s=1$, $\phi_0 = 55°$. The Stark decelerator is programmed to remove a constant kinetic energy of 0.14 cm$^{-1}$ per electric field stage from NO radicals in the $M_J\Omega=-9/4$ component of the $X\,^2\Pi_{3/2}$, v=0, J=3/2, $f$ state. A packet of NO radicals with an initial velocity of 315 m/s is selected, and decelerated to a final velocity of 253.8 m/s. The decelerated packet arrives in the detection region some 9.4 ms after production, and is split off from the remaining part of the molecular beam that is not decelerated. It is noted here that the preparation of NO radicals in the ($X\,^2\Pi_{3/2}$, v=0, J=3/2, $f$) state using excitation via a spatially well-defined laser beam results in highly structured TOF profiles in which the guided or decelerated NO radicals are well separated from the remainder of the molecular beam pulse. This facilitates the accurate interpretation of all features in the TOF profiles, and is essential for the future use of the Stark decelerated packets in, for instance, scattering experiments.

The TOF profiles that are obtained from three dimensional trajectory simulations of the experiments are shown above the experimental profiles in Figure 3. The simulated TOF profiles are in good agreement with the observed profiles for all modes of operation of the Stark decelerator. In these simulations, the individual contributions of the low-field-seeking $M_J\Omega=-3/4$ and $M_J\Omega=-9/4$ components of the $X\,^2\Pi_{3/2}$, J=3/2, $f$ state are taken into account. Both components contribute to the main central peak when the decelerator is operated at $\phi_0 = 0°$. In deceleration experiments, however, only the $M_J\Omega=-9/4$ component contributes to the decelerated peak. This is illustrated in curves (f1) and (f2) of Figure 3 that display the $M_J\Omega$ composition of the full TOF profile. The Stark shift of the $M_J\Omega=-3/4$ component is a factor three smaller than the Stark shift of the $M_J\Omega=-9/4$ component, and does not contribute to the decelerated peak.

The final velocity of the packet of NO radicals can be tuned by choosing a different phase angle $\phi_0$ for the Stark decelerator. In Figure 4, the TOF profiles are shown that are observed when the decelerator is operated at $\phi_0 = 30°$, 55° and 80° (all $s = 1$), resulting in a packet of NO radicals with a final velocity of 283.9 m/s, 253.8 m/s, and 229.2 m/s, respectively. Only the segment of the TOF profile that contains the decelerated molecules is shown. For comparison, the central peak of the TOF profile that is obtained for $s = 1$, $\phi_0 = 0°$ is shown in Figure 4 as well. The maximum intensity of the decelerated peak decreases for lower final velocities, following the decreasing acceptance of the Stark decelerator for higher values of $\phi_0$.

**Conclusion**
In this paper, we demonstrate the Stark deceleration of a molecular beam of NO radicals, which is one of the most widely used and most intensely studied species in molecular beam experiments. Due to its very small dipole moment of 0.16 D, Stark deceleration of NO radicals has long been considered unfeasible. We prepared the NO radicals in the $X\,^2\Pi_{3/2}$, v=0, J=3/2 spin-orbit excited state from the $X\,^2\Pi_{1/2}$, v=0, J=1/2 ground state by Franck-Condon pumping via the $A\,^2\Sigma^+$ state. The larger effective dipole moment in the $X\,^2\Pi_{3/2}$, v=0, J=3/2 state, in combination with a 316-stages-long Stark decelerator, allowed us to decelerate NO radicals from 315.0 m/s to 229.2 m/s, removing 47 % of their kinetic energy. The measured time-of-flight profiles of the NO radicals exiting the decelerator show good agreement with the outcome of numerical trajectory simulations. The packets of NO radicals with a tunable velocity as produced here are ideally suited for, in particular, scattering experiments of the NO radicals with atoms, molecules, and surfaces. The intensity of the packets of NO $X\ ^2\Pi_{3/2}$, v=0, J=3/2 can be further enhanced by adding an additional laser to allow for stimulated emission pumping in the preparation scheme. The tunability of the final velocity can be further increased by using decelerators of different designs. A decelerator consisting of ring-shaped electrodes [31], for instance, is better suited for molecular species with a small Stark-shift-over-mass-ratio like the NO radical. It is noted that NO radicals prepared in the $X\ ^2\Pi_{3/2}$, v=0, J=3/2 state possess a 1.2 Bohr magneton magnetic moment, making efficient deceleration using a Zeeman decelerator a good alternative.

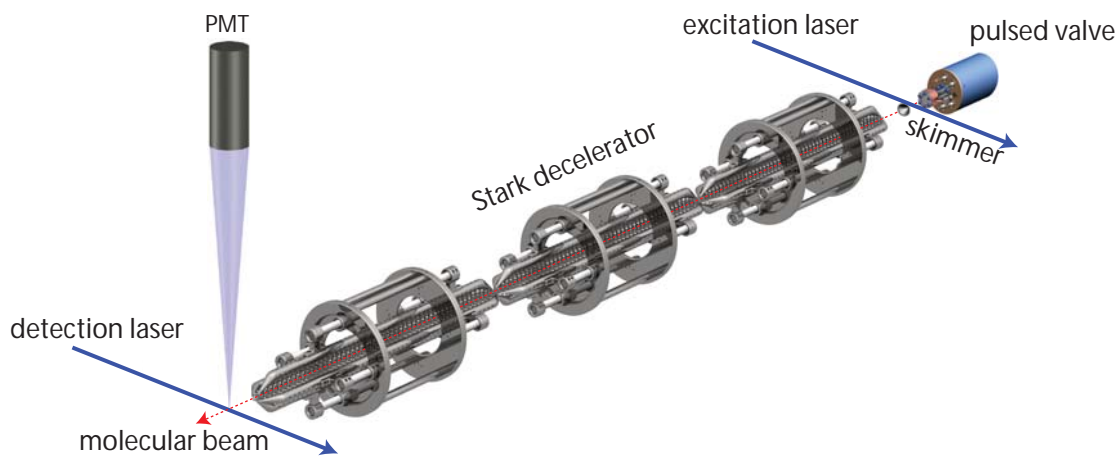

FIG. 1
Scheme of the experimental setup. A pulsed molecular beam of NO is produced by a supersonic expansion of a gas mixture (20% NO seeded in Xe) from a cooled pulsed valve. After the supersonic expansion, just in front of the skimmer, a packet of NO $X\ ^2\Pi_{3/2}$, v=0, J=3/2 radicals is created from the $X\ ^2\Pi_{1/2}$, v=0, J=1/2 ground state via Franck-Condon pumping via the $A^2\Sigma^+$ state using a pulsed dye laser. After passing through the 2.6-meter-long Stark decelerator, the NO $X\ ^2\Pi_{3/2}$, v=0, J=3/2 radicals are detected using a saturated laser induced fluorescence scheme.

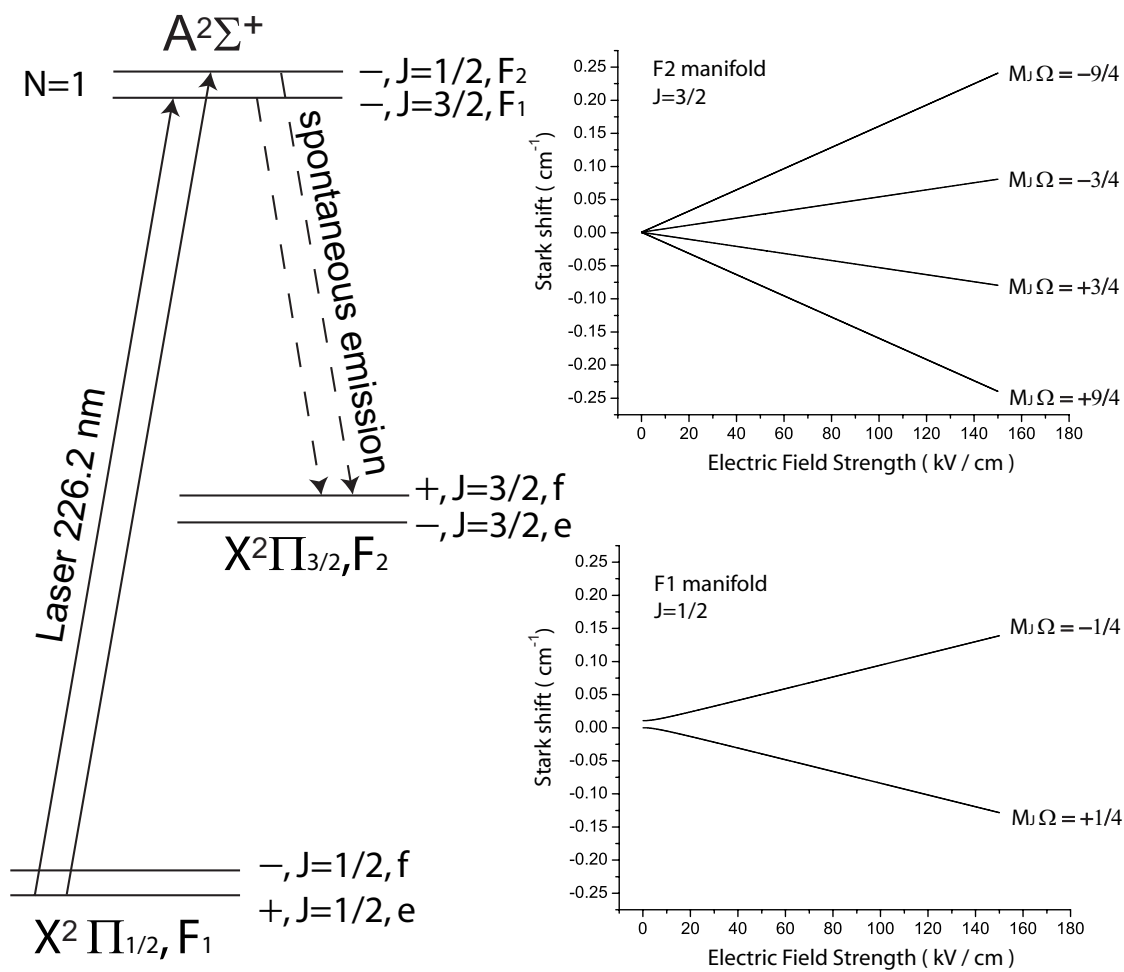

FIG. 2
Schematic representation of the energy level diagram of the NO radical, showing only the states that are of relevance to the experiments. The electronic transitions that are involved in the optical excitation scheme to produce a packet of NO radicals in the $X\ ^2\Pi_{3/2}$, v=0, J=3/2 state are shown as well. The Stark shifts for the $X\ ^2\Pi_{1/2}$, J = 1/2 and the $X\ ^2\Pi_{3/2}$, J = 3/2 states are shown in plots (b) and (c), respectively. The vertical axis indicates the energy difference relative to the field free case. The Λ-doublet splittings for the $X\ ^2\Pi_{1/2}$, J =1/2 and the $X\ ^2\Pi_{3/2}$, J =3/2 states in zero electric field amount to 0.0106 cm$^{-1}$ and 0.0013 cm$^{-1}$, respectively.

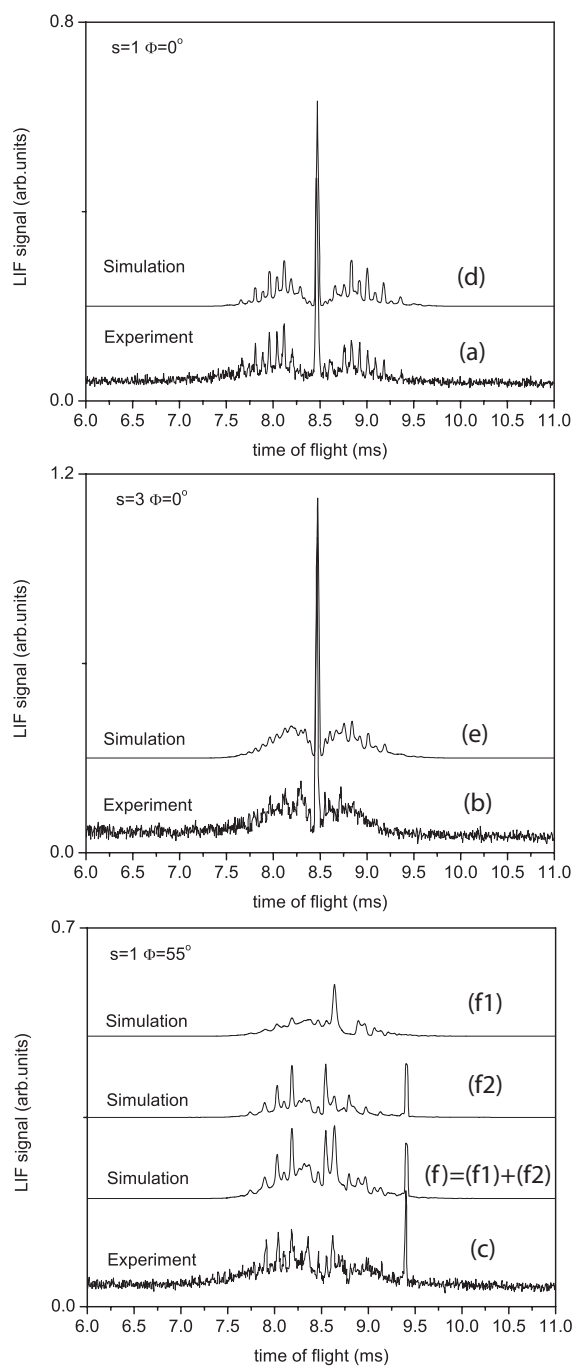

FIG. 3
Time-of-flight profiles of NO $X\ ^2\Pi_{3/2}$, v=0, J=3/2 radicals exiting the Stark decelerator when the decelerator is operated at s=1, $\phi_0 = 0°$ (curve (a)), s=3, $\phi_0 = 0°$ (curve (b)) and s=1, $\phi_0 = 55°$ (curve (c)). The time-of-flight profiles that result from three-dimensional trajectory simulations of the experiment are shown above the experimental results (curves (d), (e) and (f)). For $\phi_0 = 55°$, the individual contributions of the $M_J\Omega=-3/4$ and the $M_J\Omega=-9/4$ components to the time-of-flight profile are shown as curves (f1) and (f2), respectively. The simulated TOF profiles are given a vertical offset for clarity.

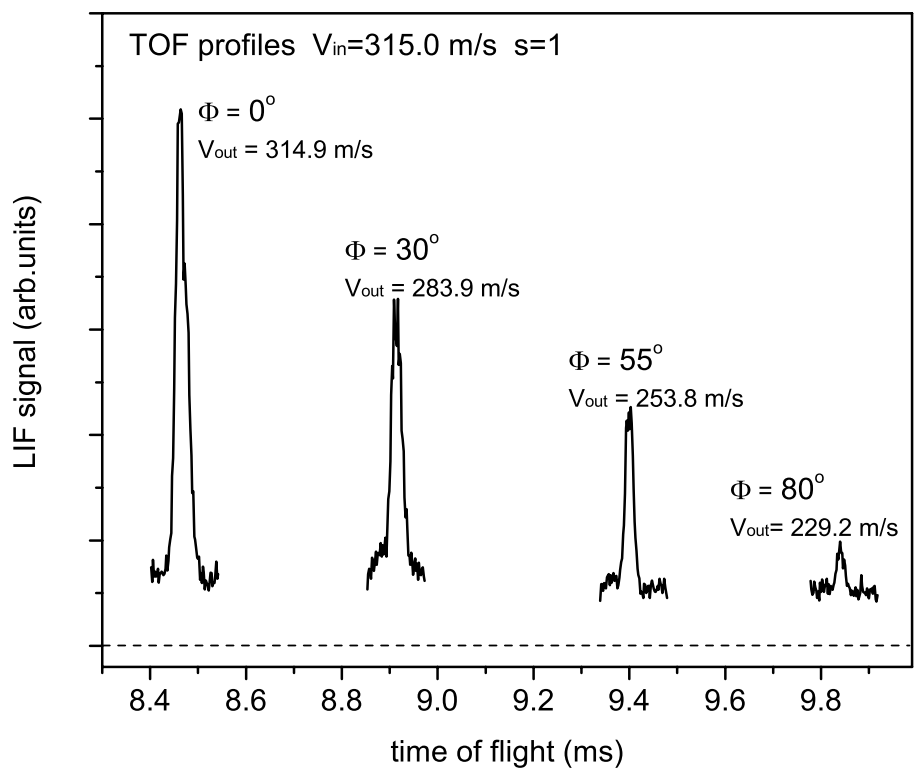

FIG. 4

Time-of-flight profiles of NO $X\,^2\Pi_{3/2}$, v=0, J=3/2 radicals exiting the Stark decelerator when the decelerator is operated at s =1 and $\phi_0$ = 0°, 30°, 55° and 80°. The time sequence was calculated for NO radicals with an initial velocity of 315.0 m/s that reside in the $M_J\Omega$=-9/4 component of the $X\,^2\Pi_{3/2}$, v=0, J=3/2 state. For these settings, the final velocities of the NO packets are 314.9 m/s, 283.9 m/s, 253.8 m/s and 229.2 m/s, respectively.

**Acknowledgments:** X.W. acknowledges the Alexander von Humboldt foundation for a research fellowship. G.M. acknowledges support from the ERC-2009-AdG under grant agreement 247142-MolChip. S.Y.T.v.d.M. acknowledges support from the Netherlands Organisation for Scientific Research (NWO) via a VIDI grant. We thank Alexander Klenner for his help to produce NO radicals in the spin-orbit excited state. We dedicate this paper to the memory of Kevin Strecker, who recently passed away. Kevin has worked for several years on the kinematic cooling of NO radicals in collaboration with David W. Chandler. We have lost a respected colleague and a dear friend.

**References:**
[1] S. Y. T. van de Meerakker, H. L. Bethlem, N. Vanhaecke, and G. Meijer, Chem. Rev. 112 (2012) 4828.
[2] H. L. Bethlem, G. Berden, F. M. H. Crompvoets, R. T. Jongma, A. J. A. van Roij, and G. Meijer, Nature 406 (2000) 491.
[3] B. C. Sawyer, B. K. Stuhl, D. Wang, M. Yeo, and J. Ye, Phys. Rev. Lett. 101 (2008) 203203.
[4] J. van Veldhoven, J. Küpper, H. L. Bethlem, B. Sartakov, A. J. A. van Roij, and G. Meijer, Eur. Phys. J. D 31 (2004) 337.
[5] E. R. Hudson, H. J. Lewandowski, B. C. Sawyer, and J. Ye, Phys. Rev. Lett. 96 (2006) 143004.
[6] H. L. Bethlem, M. Kajita, B. Sartakov, G. Meijer, and W. Ubachs., Eur. Phys. J. Spec. Top. 163 (2008) 55.
[7] J. J. Gilijamse, S. Hoekstra, S. Y. T. van de Meerakker, G. C. Groenenboom, and G. Meijer, Science 313 (2006) 1617.
[8] M. Kirste, X. Wang, H. C. Schewe, G. Meijer, K. Liu, A. van der Avoird, L. M.C. Janssen, K. B. Gubbels, G. C. Groenenboom, and S. Y. T. van de Meerakker, Science 338 (2012) 1060.
[9] A. B. Henson, S. Gersten, Y. Shagam, J. Narevicius, and E. Narevicius, Science 338 (2012) 234.
[10] L. Scharfenberg, J. Kłos, P. J. Dagdigian, M. H. Alexander, G. Meijer, and S. Y .T. van de Meerakker, Phys. Chem. Chem. Phys. 12 (2010) 10660.
[11] L. Scharfenberg, K. B. Gubbels, M. Kirste, G. C. Groenenboom, A. van der Avoird, G. Meijer, and S. Y. T. van de Meerakker, Eur. Phys. J. D 65 (2011) 189.
[12] H. L. Bethlem, G. Berden, and G. Meijer, Phys. Rev. Lett. 83 (1999) 1558.
[13] S. A. Meek, H. Conrad, and G. Meijer, Science 324 (2009) 1699.
[14] L. Scharfenberg, H. Haak, G. Meijer, and S. Y. T. van de Meerakker, Phys. Rev. A 79 (2009) 023410.
[15] S. Y. T. van de Meerakker, N. Vanhaecke, M. P. J. van der Loo, G. C. Groenenboom, and G. Meijer, Phys. Rev. Lett. 95 (2005) 013003.
[16] N. Vanhaecke, U. Meier, M. Andrist, B. H. Meier, and F. Merkt, Phys. Rev. A 75 (2007) 031402.
[17] S. D. Hogan, D. Sprecher, M. Andrist, N. Vanhaecke, and F. Merkt, Phys. Rev. A 76 (2007) 023412.
[18] E. Narevicius, A. Libson, C. G. Parthey, I. Chavez, J. Narevicius, U. Even, and M. G. Raizen, Phys. Rev. Lett. 100 (2008) 093003.
[19] A. Trimeche, M. N. Bera, J. -P. Cromières, J. Robert, and N. Vanhaecke, Eur. Phys. J. D 65 (2011) 263.
[20] E. Narevicius, A. Libson, C. G. Parthey, I. Chavez, J. Narevicius, U. Even, and M. G. Raizen, Phys. Rev. A 77 (2008) 051401.

[21] E. Narevicius and M. G. Raizen, Chem. Rev. 112 (2012) 4879.

[22] H. Kohguchi, T. Suzuki, and M. H. Alexander, Science 294 (2001) 832.

[23]  C. J. Eyles, M. Brouard, C. -H. Yang, J. Kłos, F. J. Aoiz, A. Gijsbertsen, A. E. Wiskerke, and S. Stolte, Nature Chem. 3 (2011) 597.

[24] N. H. Nahler, J. D. White, J. LaRue, D. J. Auerbach, and A. M. Wodtke, Science 321 (2008) 1191.

[25] The effective dipole moment is given by $\mu M\Omega/J(J+1)$, where M and $\Omega$ are the projection of J onto the space-fixed quantization axis and the internuclear axis, respectively.

[26] M. S. Elioff, J. J. Valentini, and D. W. Chandler, Science 302 (2003) 1940.

[27] R. Fulton, A. I. Bishop, M. N. Shneider, and P. F. Barker, Nat. Phys. 2 (2006) 465.

[28] J. Brzozowski, N. Elander, and P. Erman, Physica Scripta 9 (1974) 99.

[29] S. Y. T. van de Meerakker, N. Vanhaecke, H. L. Bethlem, and G. Meijer, Phys. Rev. A 71 (2005) 053409.

[30] L. Scharfenberg, S. Y. T. van de Meerakker and G. Meijer, Phys. Chem. Chem. Phys., 13, (2011) 8448.

[31] S. A. Meek, M. F. Parsons,G. Heyne, V. Platschkowski, H. Haak, G. Meijer, and A. Osterwalder, Rev. Sci. Instrum. 82 (2011) 093108.